\documentclass[aps,twocolumn,prl,showpacs,preprintnumbers,amsmath,amssymb]{revtex4-1}
\newcommand{\ket}[1]{\left|#1\right>}

\usepackage{graphicx}
\usepackage{dcolumn}
\usepackage{bm}
\usepackage{amsmath}
\usepackage{times}
\usepackage[usenames]{color}
\usepackage{natbib}
\usepackage{ulem}

\begin{document}
\title{Photoinduced inhomogeneous melting of charge order by ultrashort pulsed light excitation}
\author{Hitoshi Seo$^{1,2}$ and Yasuhiro Tanaka$^{3}$}
\affiliation{$^1$Condensed Matter Theory Laboratory, RIKEN, Wako 351-0198, Japan}
\affiliation{$^2$Center for Emergent Matter Science (CEMS), RIKEN, Wako 351-0198, Japan}
\affiliation{$^3$Kanazawa Institute of Technology, Kanazawa 921-8501, Japan}
\date{\today}
\begin{abstract} 
We numerically investigate photo-responses of the charge ordered state upon stimuli of pulsed laser light, 
 especially paying attention to the differences in the pulse width,  whose shortness has been a key to experimentally realize large photo-induced effects. 
As a 
model for charge ordering, we consider an interacting spinless fermion model on a one-dimensional chain coupled to classical phonons 
and numerically simulate its real-time dynamics. 
First, we demonstrate the case where spatially uniform dynamics take place, 
 and show the delayed response of phonons after the photo-irradiation causing destabilization of charge order. 
When a local impurity potential is introduced in the initial state, 
  an inhomogeneous charge order melting occurs and the delay in the time scale also happens in the real space melting process. 
Finally, by investigating wider parameter ranges using different pulse widths, 
 we find that the use of ultrashort pulses considerably expands the frequency window for charge order melting; 
 it is indispensable in realizing the photo-induced phase transition with off-resonant conditions. 
\end{abstract} 
%
%
%
%

\maketitle

{\it Introduction---.}~Photo-responses of condensed matter are increasingly upon intense research 
 thanks to rapid advancements in technologies~\cite{Orenstein_PhysToday2012,Kampfrath_NatPhoto2013,delaTorre_RMP2021}. 
Among the rich non-thermal phenomena caused by pulsed laser light in solid states through electron-light interaction, 
 optically switching one phase to another, i.e., photo-induced phase transition (PIPT), 
 is a classical and still intensively-studied theme~\cite{Nasu_book2004,Sato_AccChemRes2003,Koshihara_PhysRep2022}. 
By  investigating the initial processes of PIPTs in different materials
 taking advantage of the modern techniques achieving high time resolution,  
 experimental works revealed that the response of the system subsequent to the light absorption often happens in a ultrafast time scale, 
 such as in femtosecond order~\cite{Iwai_JPSJ2006,Iwai_Crystals2012,Nicoletti_AOP2016,Gianetti_AdvPhys2016}. 
This indeed corresponds to the energy scale of the Fermi degenerate many-electron states in typical solids. 
Therefore, to explore the mechanism of PIPT, understanding the initial process of the interacting electron system is crucial. 

To theoretically investigate PIPTs within microscopic models that describe many-body system subject to laser light, 
 mainly two different approaches have been adopted so far. 
One is to assume continuous-wave light where we can take advantage of the Floquet theory for time-periodic systems~\cite{Aoki_RMP2014,Oka_AnnuRev2019}.  
The other is to numerically simulate the time evolution directly introducing the pulsed light~\cite{Yonemitsu_Crystals2012,Ishihara_JPSJ2019}. 
In any case, 
 the pump light excites the system so that its absorption disturbs the ground state which may evolve into a non-equilibrium state. 
Therefore the light frequency is usually chosen so that the absorption is large, 
 whose range can be deduced from the linear optical spectrum at equilibrium. 
Typical example is to use frequencies across the one-particle gap to make 
 photo-doping~\cite{Takahashi_PRL2002,Murakami_CommPhys2022}, 
 or, more efficiently, at resonant conditions inducing 
 collective excitations~\cite{Kawakami_PRL2010,Matsunaga_Science2014,Hashimoto_JPSJ2014,Hashimoto_JPSJ2015}. 

However, in the actual experiments, different light frequencies are practically chosen by technical reasons; 
 there the pulsed light frequency is sometimes even far away from the large absorption range, 
 but still in many cases PIPTs are observed~\cite{Iwai_PRL2003,Iwai_PRL2007,Kawakami_PRL2010,Onda_PRL2008,Rohwer_Nat2011,Mohr-Vorobeva_PRL2011,Porer_NatMat2014,Itoh_PRR2021}.  
 This raises a fundamental problem about how we can theoretically describe PIPTs without fine tuning of the light frequency. 
 One key ingredient to reconcile this discrepancy is to use ultrashort pulsed light that excites the system over a wide frequency window. 
 It also reduces the heating effect and then has advantages in the search for possible hidden states of matter~\cite{Fiebig_ApplPhysB2000,Fausti_Science2011,Ichikawa_NatMater2011,Stojchevska_Science2014,Ishikawa_NatComm2014}.  
Another important factor is the emergent inhomogeneity during PIPT processes. 
In some materials it has been revealed that the real space expansion of the competing phase is a key for
 PIPT~\cite{Nagaosa_PRB1989,Iwai_PRL2002,Miyashita_JPSJ2003}: the so-called domino effect. 
In fact, in our previous work~\cite{Seo_PRB2018}, 
 using a simplest model for charge-order (CO) insulator, i.e., an interacting spinless fermion model, 
 we found that an inhomogeneous melting of CO can be triggered by a kink structure in the initial state. Importantly, 
 the initial inhomogeneity (the kink) embedded in the bulk CO is essential for the appearance of photoinduced metallic 
 state suggesting its vital role in describing PIPT. 

These facts indicate that, to understand the mechanism of PIPT, a systematic investigation on the role of light frequency,  pulse width, and inhomogeneity is highly desirable. 
Specifically, this is indispensable to the optimization of conditions to make PIPTs happen, and to the efficient photocontrol of materials.  
Here we develop the theory to a general situation including electron-phonon interaction, 
 which brings about different time scales, 
 and taking account of an impurity potential as the initial inhomogeneity. 
By simulating the real-time evolution of the electron-light coupled model considering different pulse widths,  
 we find that ultrashort pulsed light can induce the inhomogeneous CO melting 
 even when the frequency in the oscillatory part is completely off the resonant condition. 

{\it Model and method---.} We treat an interacting spinless fermion model on a one-dimensional chain with Holstein-type coupling to classical phonons, 
whose Hamiltonian is given as 
\begin{align}
\mathcal{H}=
&\sum_{i}  \left( t\ c_{i}^{\dagger}c_{i+1} + t'\ c_{i}^{\dagger}c_{i+2}+\text{h.c.}\right) + \sum_{i}V n_{i} n_{i+1} \notag\\
 & - g \sum_{i} q_i n_i + \frac{\omega_\textrm{lat}}{2} \sum_{i} \left( q_i ^2 + p_i ^2 \right), 
\label{eqn1}
\end{align}
where $c_{i} (c_{i}^{\dagger})$ is the annihilation (creation) operator of a spinless fermion at site $i$,
and $n_{i} \equiv c_{i}^{\dagger}c_{i}$ is the fermion number operator. 
$t$ and $t'$ are the hopping integrals between site pairs of nearest neighbor and next nearest neighbor, 
 and $V$ is the nearest neighbor Coulomb repulsion which induces CO at half-filling.  
The effect of external laser pulse polarized along the chain direction 
 is incorporated by the Peierls substitution in the hopping integrals  
 with a time ($\tau$) dependent gauge field $A(\tau)$ as, 
$t \rightarrow t(\tau)= t\ e^{iA(\tau)}, \hspace{2mm} t' \rightarrow t'(\tau)=t' e^{2iA(\tau)}$,
where the lattice constant, the light velocity, the elementary charge, and $\hbar$ 
are taken as unity.  
The pump light $A(\tau)$ is centered at $\tau=0$ 
 with a gaussian envelope and frequency $\omega_{p}$, written as  
\begin{align}
A(\tau)=\frac{A_{p}}{\sqrt{2\pi}\tau_{p}}\exp{\left(-\frac{\tau^{2}}{2\tau_{p}^{2}}\right)}\cos(\omega_{p}\tau),
\label{eqn2}
\end{align}
where $A_{p}$ and $\tau_{p}$ control the amplitude and the pulse width, respectively. 
Note that, in this form, the time integral of the gaussian part does not depend on $\tau_p$. 
The phonon coordinates $q_i$ and momenta $p_i = \dot{q_i} / \omega_\textrm{lat}$ are time-dependent classical values. 
In the following we set $t=1$ as the unit of energy (and time $1/t$), and fix $t'=0.1$ and $V=2$; 
 electron-phonon parameters are taken as  $g=0.2$ and $\omega_\textrm{lat} = 0.1$.  

For the time evolution, we combine the time-dependent Hartree-Fock method 
 at absolute zero temperature~\cite{Terai_PTP1993,Kuwabara_JPSJ1995,Tanaka_JPSJ2010}  
 for the electronic part and assume the Hellmann-Feynman theorem at each time step 
 to calculate the force entering the Newtonian equation for the phonon variables. 
As formulated in Refs.~\cite{Terai_PTP1993,Kuwabara_JPSJ1995,Tanaka_JPSJ2010,Hashimoto_JPSJ2015}, 
 the wave function at time $\tau$,  $\ket{\Psi(\tau)}$, 
 is calculated using the discretized Schr$\rm \ddot o$dinger equation, 
 and the leap-frog method is used for solving the equation of motion for phonons. 
Note that, even though we are treating a one-dimensional model~\cite{Hashimoto_PRB2017}, 
 by using Hartree-Fock approximation and classical treatment of phonons, 
 we are implying higher dimension (quasi-one-dimensional) systems. 

In the following, we consider half-filling and the periodic boundary condition. 
When a system size $N$ is even in Eq.~(\ref{eqn1}), 
 starting from the ground-state Slater determinant describing the CO insulating state
 in which charge rich and poor sites arrange alternatively, 
 the system shows coherent dynamics without any inhomogeneity. On the other hand, when we choose an odd $N$, 
 the lowest-energy initial state contains a kink structure between CO domains
 and inhomogeneous dynamics is seen. 
These are similar to our previous results for a purely electronic model~\cite{Seo_PRB2018}. 
For the initial inhomogeneity, instead of using odd $N$ with the kink structure, 
 we treat a system with a single impurity added, described by $\mathcal{H} + u_\textrm{imp} \ n_{i_\textrm{imp}}$, 
 where $u_\textrm{imp}$ is a local impurity potential at site $i_\textrm{imp}$. 
We find that the impurity indeed initiates the CO melting similarly to the kink structure, and  
 here we only show results for even $N$ with and without the impurity. 
 We choose $N=1000$, which is large enough to neglect finite-size effect within the time domain we consider, 
 typically $\tau \leq 150$. The energy gap in the initial state is $\Delta_\textrm{CO} = 2.59$. 

\begin{figure}[tb]
\begin{center}
\includegraphics[width=8cm]{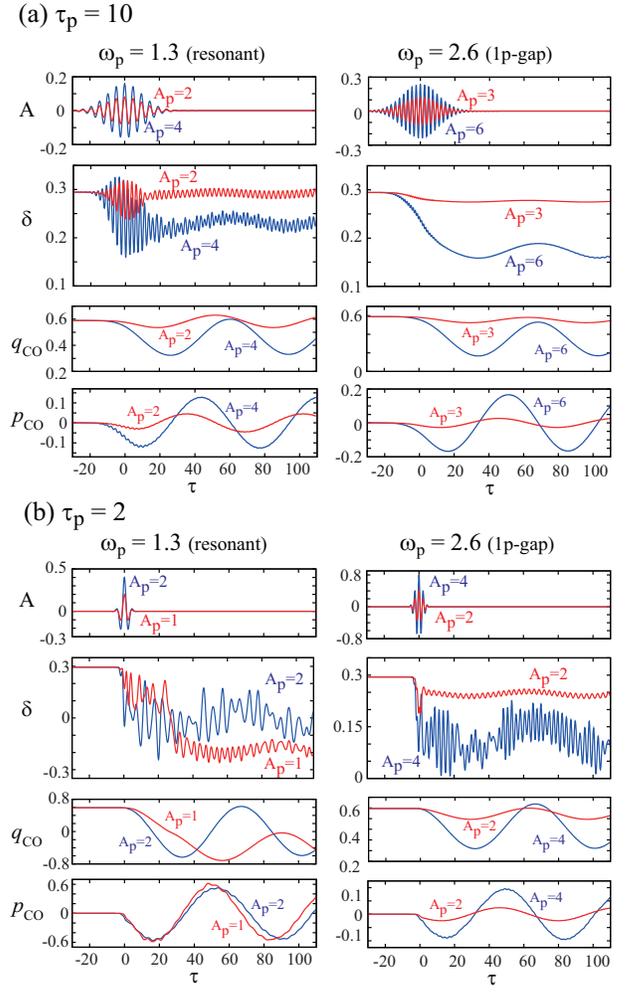}
\end{center}
\vspace{-0.3cm}
\caption{(Color online) $\tau$-dependence of the CO order parameter $\delta$ 
 and the two-fold component of the phonon coordinates 
  $q_\textrm{CO}$ and momentum  $p_\textrm{CO}$, 
 for widths of the pump pulse taken as (a) moderate $\tau_p=10 $ and (b) ultrashort $\tau_p=2$, 
 whose form of gauge field $A(\tau)$ is also plotted.  
Cases with two pump-light frequencies, 
 $\omega_p = 1.3$ (left) and $\omega_p = 2.6$ (right) are chosen, 
 for different intensities $A_p$ shown in the Figure. 
}
\label{fig1}
\end{figure}

{\it Uniform dynamics---.}~
Without the impurity potential, 
 the two-fold periodicity of charge rich and poor sites is maintained throughout the simulation, namely, coherent dynamics of CO are seen. 
The expectation values of the charge density, the phonon coordinates, and momenta are described as  
 $\langle n_i \rangle = 1/2 + (-1)^i \delta(\tau)$, $q_i = (-1)^i q_\textrm{CO}(\tau)$,  and $p_i = (-1)^i p_\textrm{CO}(\tau)$, respectively; 
 here, $\delta(\tau)$ is the time-dependent CO order parameter. 
In Fig.~\ref{fig1}, 
 we show the $\tau$-dependence of  $\delta(\tau)$, $q_\textrm{CO}(\tau)$, $p_\textrm{CO}(\tau)$,  
 together with the gauge field $A(\tau)$, 
 for two choices of pump pulse widths, (a) $\tau_p=10$ (moderate) and  (b) $\tau_p=2$ (ultrashort). 
The cases for pump-light frequencies $\omega_p = 1.3$ (left panels) and $\omega_p =2.6$ (right panels)
 are plotted, approximately corresponding to the resonant charge transfer energy and the one-particle energy gap $\Delta_\textrm{CO}$; 
 we call them resonant and 1p-gap 
 conditions, respectively, in the following.  

In the case of a moderate width pulse of $\tau_p=10$ in Fig.~\ref{fig1}(a), 
 for both resonant and 1p-gap 
 conditions, coherent motion of phonons is induced following the charge response. 
Then, the degree of CO, $\delta(\tau)$, in turn follows the phonons 
 so that the charge degree of freedom also keeps oscillating at the phonon frequency $\omega_\textrm{lat}$. 
This is in contrast to the purely electronic model where $\delta(\tau)$
 stays constant after the light is switched off~\cite{Seo_PRB2018}. 
The notable difference in the two pump frequencies is the presence of a fast oscillation in $\delta(\tau)$ 
 seen only in the resonant condition, 
 which is a direct response to the pump light with frequency $\omega_p$; 
 the nature of this resonant mode is discussed in detail in our previous paper~\cite{Seo_PRB2018}. 

On the other hand, for the ultrashort pulse of $\tau_p=2$, displayed in Fig.~\ref{fig1}(b), 
 such $\omega_p$ oscillations are prominent not only in the resonant but also in the 1p-gap 
 conditions.  
The response itself is large, if we compare with the case of $\tau_p=10$ for the same value of $A_p$. 
We should note that, in the functional form in Eq.~(\ref{eqn2}), 
 the maximum value in $A(\tau)$ becomes higher for shorter pulses for fixed $A_p$ (see top panels in Fig.~\ref{fig1})
 since the integral of the pulse intensity is constant, as mentioned above. 

{\it Inhomogeneous CO melting---.}~
Next we show results for the system with a single impurity site. 
The data below are for the case of $u_\textrm{imp}=0.1$, 
 but most of the results presented here are 
 actually insensitive to the value of $u_\textrm{imp}$, irrespective to its sign, 
 even though the details of the simulation depend on it; 
 actually the results are very similar to case when the initial state has a kink structure for $N$ odd systems. 
In this sense, the impurity acts as a seed for the emergence of inhomogeneity. 
 
\begin{figure}[tb]
\begin{center}
\includegraphics[width=8cm]{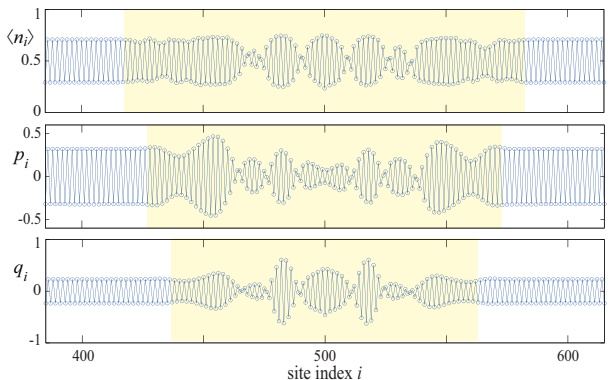}
\end{center}
\vspace{-0.3cm}
\caption{(Color online) 
Example of the inhomogeneous real space pattern of charge density $\langle n_i \rangle$, 
 phonon momentum $p_i$, and its coordinate $q_i$.
The parameters of the pump pulse are $\tau_p=2$ and $A_p=1$; the pattern at time $\tau=35$
is plotted for sites $i=375$ -- $625$ within the system of $N=1000$ with the impurity potential at site $i_\textrm{imp}=500$.
The inhomogeneity in each variable is shown by the shaded region as a guide to the eye.}
\label{fig2}
\end{figure}
In Fig.~\ref{fig2}, we show an example of how the CO melting occurs in an inhomogeneous way, for the ultrashort pulse light ($\tau_p=2$); 
 the real space image of the melting process does not depend much on the pulse width.  
The disordered area are expanding from the impurity site (at $i_\textrm{imp}=500$)  embedded in the bulk CO state, 
 whose snapshot for $\tau=35$, well after the pump pulse,  is shown in the figure. 
The inhomogeneous region, represented by the shaded region in each variable, 
 $\langle n_i \rangle$, $p_i$, and $q_i$, expands as time evolves. 
Importantly, 
 one can see a time delay in the phonon variables compared to the charge degree of freedom 
 in terms of the disordered regions in real space. 
Microscopically, 
 first the CO melting occurs starting from the neighborhood of the impurity site, 
 and then, through the Newton equation coupling the momemtum $p_i$ and 
 the Hellmann-Feynman force which contains the charge density $\langle n_i \rangle$, 
 the disorder in charge brings about inhomogeneity in the spacial distribution of momemtum $p_i$. 
Finally,  it affects the  phonon coordinate $q_i$, whose time dependence is given by $\dot{q_i} = \omega_\textrm{lat} \ p_i$. 

\begin{figure}[tb]
\begin{center}
\includegraphics[width=8cm]{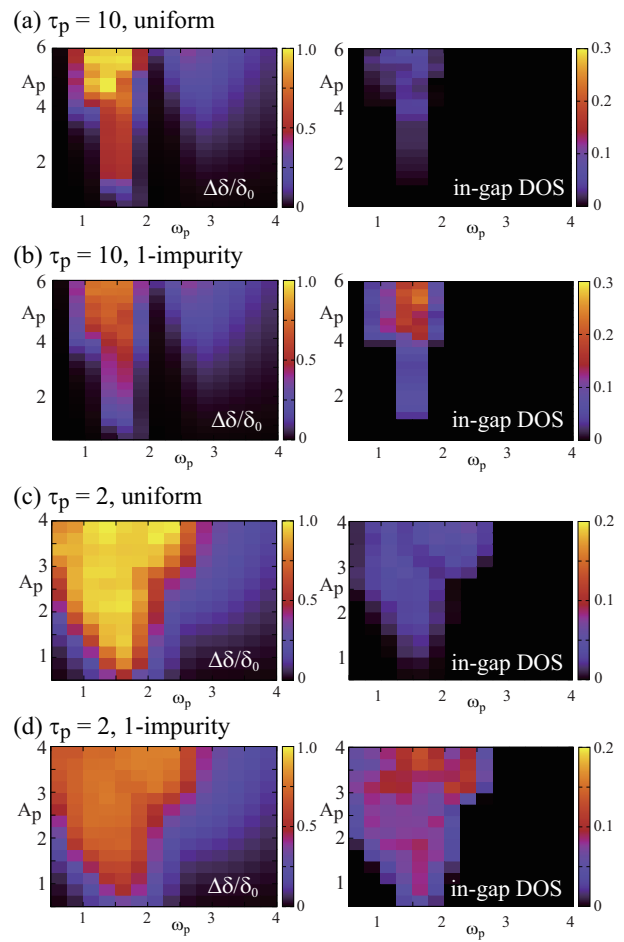}
\end{center}
\vspace{-0.5cm}
\caption{(Color online) 
Two-dimensional plots on the \{$\omega_p, A_p$\} plane of the time averaged photo-responses: 
In the left panels, decrease in the order parameter $\Delta\delta$ 
 normalized by its equilibrium value $\delta_0$, 
 and in the right panels, in-gap density of states (DOS) are plotted. 
For their definition, see text. 
The pump pulse width is $\tau_p=10$ [$\tau_p=2$], 
 for (a) [(c)] the uniform case without the impurity potential and 
 (b) [(d)] the inhomogeneous case with the impurity potential $u_\textrm{imp}=0.1$.
}
\label{fig3}
\end{figure}
To seek for the conditions for the CO melting to occur 
 in a broader parameter range, in Fig.~\ref{fig3},
 we show two-dimensional plots on the \{$\omega_p, A_p$\} plane
 of the decrease in CO order parameter $\delta$ and the in-gap density of states (DOS) as a measure of CO melting, 
 for different situations. 
The decrease in $\delta$, $\Delta \delta$, which is averaged over time $\tau = 20$ to 150, 
 is normalized by the initial value $\delta_0 = 0.295$; 
 for the inhomogeneous cases we take the real-space average for the whole system for each time step. 
The in-gap DOS 
 is defined as the DOS within the energy range $\pm 0.1$ from the Fermi energy 
 taken as the midpoint between the $(N/2)$-th and $(N/2+1)$-th eigenenergies; 
 it is normalized by the value for the metallic state for $V=0$. 
 This in-gap DOS is a measure of the partial DOS well inside the initial gap ($\Delta_\textrm{CO} = 2.59$), 
 sensing the real-space CO melting~\cite{Seo_PRB2018}. 

For a moderate pump pulse width of $\tau_p = 10$, the results without and with the impurity potential are shown 
 in Figs.~\ref{fig3}(a) and  \ref{fig3}(b), respectively. 
The CO melting is realized only near the resonant conditions of $\omega_p\sim 1.5$, in the presence of impurity [Fig.~\ref{fig3}(b)]. 
The area with finite but small in-gap DOS seen for the uniform case in Fig.~\ref{fig3}(a) suggests 
 CO phase inversions occurring instantaneously during the time evolution; 
 this is basically the same area where the inhomogeneous CO melting with large in-gap DOS is observed in the presence of impurity~\cite{Seo_PRB2018}.
The response for the charge across the gap $\Delta_\textrm{CO}$ is only seen in  $\Delta \delta/\delta_0$, 
 for both Figs.~\ref{fig3}(a) and \ref{fig3}(b). 
These features are similar to the purely electronic case in Ref.~\cite{Seo_PRB2018} 
 where we considered a kink structure as the initial inhomogeneity. 

In the case for the ultrashort pulse with $\tau_p = 2$ in Figs.~\ref{fig3}(c) and \ref{fig3}(d), 
 without and with the impurity potential, respectively, 
 one can see the considerably enlarged area of large response, up to about $\omega_p\sim 2.5$. 
As indicated by the in-gap DOS plotted in Fig.~\ref{fig3}(d), 
 CO melting can occur in this broad range of frequency, which is in sharp contrast to the case of $\tau_p=10$ 
 [Fig.~\ref{fig3}(b)]. 
The reason for such an expansion is most likely a wider range of frequency 
 effectively exciting the system by making the pulse width shorter; 
 by Fourier transformation of $A(\tau)$ in Eq.~(\ref{eqn2}), the shorter width in time, 
 the pulse contains the wider window in energy.  
We note that, in the moderate pump pulse cases, 
 the inclusion of the electron-phonon interaction does not change the behavior of CO melting overall, 
 compared to the case without it. 
In contrast, when we use the ultrashort pulse, in the purely electronic case the response is affected 
 by  the small number of cycles contained in the pulse [see $A(\tau)$ in Fig.~\ref{fig1}], 
 resulting in a very irregular behavior in the \{$\omega_p, A_p$\}  dependence, 
 presented in the Supplemental Material~\cite{Supplement}. 
Such an irregularity is blurred in the case here with the Holstein-type coupling, 
which plays a role as a heat bath with respect to the electronic system and makes different fluctuations mixed. 
In this sense, the electron-phonon interaction acts in favor of realizing the CO melting under general conditions in the pump pulse.

{\it Conclusion---.}~ 
In this work, we simulated the time evolution of a CO system, 
 modeled by an interacting spinless fermions coupled to classical phonons. 
Different time scales between the charges interacting via electron-electron Coulomb repulsion and the electron-phonon interaction 
 bring about delay of  phonons not only in time dependence of their responses 
 but also in the real-space inhomogeneous CO melting process. 
We found that the inhomogeneous CO melting out of a homogeneous light emission,  
 through the cooperative effect of resonant excitation and the triggering by small disorder (impurity discussed here), 
 can be realized even well off this resonant condition, by the use of ultrashort pulses as the pump light. 
This is indeed relevant to experiments as we mentioned in the introduction: 
 we can make PIPT by using different $\omega_p$ for the pulsed light.  
For $\omega_p$ off the resonant condition, a non-linear $A_p$ dependence is expected, 
 also seen in experiments; for long pulses, and off course for CW  as the long width limit, 
 we do not expect such an effect so a fine tuning of frequency is needed, 
 therefore the importance of ultrashort pulses in the realization of PIPT is now depicted.  


\begin{acknowledgments}
The authors would like to thank late Sumio Ishihara for collaboration in the initial stage of this work. 
This work was supported by JSPS KAKENHI Grants
Numbers 26400377, 15H02100, 16H02393, 17H02916, and 18H05208. 
\end{acknowledgments}
%


\end{document}